\documentclass[twocolumn,showpacs,preprintnumbers,amsmath,amssymb,floatfix]{revtex4-2}
\usepackage{graphicx}
\usepackage{dcolumn}
\usepackage{bm}
\usepackage{latexsym}
\usepackage[hang,small,bf]{caption}
\usepackage[subrefformat=parens]{subcaption}
\usepackage{braket}

\usepackage{here}

\usepackage{multirow} 

\usepackage{comment}

\def\Tr{\text{Tr}} 
\def\nn{\nonumber}
\newcommand{\beqn}{\begin{eqnarray}}
\newcommand{\beq}{\begin{equation}}
\newcommand{\eeqn}{\end{eqnarray}}
\newcommand{\eeq}{\end{equation}}

\newcommand{\tr}{\mathop{\rm Tr}}
\newcommand{\F}{\phantom {1}}

\newcommand{\kbra}[1] { \left< #1 \right>}

\newcommand{\rcnp}{\affiliation{Research Center for Nuclear Physics, Osaka University, Osaka 567-0047, Japan}}
\newcommand{\Kochi}{\affiliation{Library and Information Technology, Kochi University, Kochi 780-8520, Japan}}
\newcommand{\NYCU}{\affiliation{Institute of Physics, National Yang Ming Chiao Tung University, Hsinchu 30010, Taiwan}}
\newcommand{\Gkochi}{\affiliation{Department of Mathematics and Physics, Kochi University, 2-5-1 Akebono-cho, Kochi 780-8520, Japan}}

\begin{document}
\title{Monopoles of the Dirac type and color confinement in QCD \\
- First results of $SU(3)$ numerical simulations without gauge fixing -}
\author{Katsuya Ishiguro}
\email[e-mail:]{ishiguro@kochi-u.ac.jp}
\Kochi
\author{Atsuki Hiraguchi}
\email[e-mail:]{a.hiraguchi@nycu.edu.tw}
\NYCU
\Gkochi
\author{Tsuneo Suzuki}
\email[e-mail:]{tsuneo@rcnp.osaka-u.ac.jp}
\rcnp

\date{\today}

\begin{abstract}
If non-Abelian gauge fields in $SU(3)$ QCD have a line-singularity leading to non-commutativity with respect to successive partial-derivative operations, the non-Abelian Bianchi identity is violated. The violation as an operator is shown to be equivalent to violation of Abelian-like Bianchi identities. Then there appear eight Abelian-like conserved magnetic monopole currents of the Dirac type in $SU(3)$ QCD. Exact Abelian (but kinematical) symmetries appear in non-Abelian $SU(3)$ QCD. 
Here we try to show, using lattice Monte Carlo simulations of $SU(3)$ QCD, the Abelian dual Meissner effect due to the above Abelian-like monopoles are responsible for color confinement in $SU(3)$ QCD. If this picture is correct, the string tension of non-Abelian Wilson loops is reproduced fully by that of the Abelian Wilson loops. This is called as perfect Abelian dominance. Furthermore, since the linear potential in Abelian Wilson loops is caused by the solenoidal monopole currents, the Abelian string tension is fully reproduced by that of Abelian monopole potentials. It is called as perfect monopole dominance.
In this report, the perfect Abelian dominance is shown to exist with the help of the multilevel method but without introducing additional smoothing techniques like partial gauge fixings
, although lattice sizes studied are not large enough to study the infinite volume limit. 
Perfect monopole dominance on $24^3\times 4$ at $\beta=5.6$ is also shown without any additional gauge fixing but with a million of thermalized configurations.  The dual Meissner effect around a pair of static quark and antiquark is studied also on the same lattice.  
Abelian electric fields are squeezed due to solenoidal monopole currents and the penetration length for an Abelian electric field of a single color is the same as that of non-Abelian electric field. The coherence length is also measured directly through the correlation of the monopole density and the Polyakov loop pair.
The Ginzburg-Landau parameter indicates that the vacuum type is the weak type I (dual) superconductor. 
Although the scaling and the infinite-volume limits are not studied yet, the results obtained above without any additional assumptions as well as more clear previous $SU(2)$ results seem to suggest strongly the above Abelian dual Meissner picture of color confinement mechanism.   
\end{abstract}


\maketitle
\section{Introduction}

Color confinement in quantum chromodynamics (QCD) is still an important unsolved problem. 
As a picture of color confinement, 't~Hooft~\cite{tHooft:1975pu} and Mandelstam~\cite{Mandelstam:1974pi} conjectured that the QCD vacuum is a kind of a magnetic superconducting state caused by condensation of magnetic monopoles and  an effect dual to the Meissner effect works to confine color charges. 
However, in contrast to SUSY QCD~\cite{Seiberg:1994rs} or Georgi-Glashow 
model~\cite{'tHooft:1974qc,Polyakov:1976fu} with scalar fields,
to find color magnetic monopoles which
condense is not straightforward in QCD. 
If the dual Meissner effect picture is correct, it is absolutely necessary to find color-magnetic monopoles only from gluon dynamics of QCD. 

An interesting idea to introduce such an Abelian monopole in QCD is to project QCD to the Abelian maximal torus group by a partial (but singular) gauge fixing~\cite{tHooft:1981ht}.  In $SU(3)$ QCD, the maximal torus group is  Abelian $U(1)^2$. Then Abelian magnetic monopoles appear as a topological object at the space-time points corresponding to the singularity of the gauge-fixing matrix. Condensation of the monopoles  causes  the dual Meissner effect with respect to $U(1)^2$. Numerically, an Abelian projection in various gauges such as the maximally Abelian (MA) gauge~\cite{Kronfeld:1987ri,Kronfeld:1987vd} seems to support the conjecture~\cite{Suzuki:1992rw, Chernodub:1997ay}.  
Although numerically interesting, the idea of Abelian projection~\cite{tHooft:1981ht} is theoretically very unsatisfactory. Especially there are infinite ways of such a partial gauge-fixing and whether the 't Hooft scheme depends on gauge choice or not is not known. 

In 2010 Bonati et al.~\cite{Bonati:2010tz} found a relation that the violation of non-Abelian Bianchi identity (VNABI) exists behind the Abelian projection scenario in various gauges. 
Under the stimulus of the above work~\cite{Bonati:2010tz}, one of the authors (T.S.)~\cite{Suzuki:2014wya} found a more general relation that VNABI is just equal to the violation of Abelian-like Bianchi identities corresponding to the existence of Abelian-like monopoles. The Abelian-like monopole currents satisfy an Abelian conservation rule kinematically. There can exist  
exact Abelian (but kinematical) symmetries in non-Abelian QCD.  A partial gauge-fixing is not necessary at all from the beginning. If the non-Abelian Bianchi identity is broken in QCD, Abelian-like monopoles necessarily appear due to a line-like singularity leading to a non-commutativity with respect to successive partial derivatives. This is hence an extension of the Dirac idea~\cite{Dirac:1931} of monopoles in Abelian QED to non-Abelian QCD. 

In the framework of simpler $SU(2)$ QCD, some interesting numerical results were obtained. Abelian and monopole dominances as well as the Abelian dual Meissner effect are seen clearly without any additional gauge-fixing  already in 2009~\cite{Suzuki:2007jp,Suzuki:2009xy}.  But at that time, no theoretical explanation was clarified with respect to Abelian-like monopoles without any gauge-fixing. They are now found to be just Abelian-like monopoles proposed in the above paper~\cite{Suzuki:2014wya}.  Also, the existence of the continuum limit of this new kind of Abelian-like monopoles was discussed with the help of the block-spin renormalization group concerning the Abelian-like monopoles.  The beautiful scaling behaviors showing the existence of the continuum limit are observed with respect to the monopole density~\cite{Suzuki:2017lco} and the infrared effective monopole action~\cite{Suzuki:2017zdh}. 
The scaling behaviors are also independent of smooth gauges adopted.

Here it is important to note that our confinement picture~\cite{Suzuki:2014wya,Suzuki:2017lco} is completely different from the Abelian projection scheme~\cite{tHooft:1981ht} and the interpretation proposed in Ref.\cite{Bonati:2010tz}. Bonati et al. say that gauge invariance of various 't Hooft Abelian projections is proved directly with the help of VNABI. Their statements in contradiction to ours~\cite{Suzuki:2014wya,Suzuki:2017lco} are not however justified as explicitly shown in a separate work done by one of the authors (T.S.)~\cite{Suzuki:20220422}. 

It is very interesting to study the new Abelian-like monopoles in $SU(3)$ QCD.  To check if the Dirac-type monopoles are a key quantity of color confinement in the continuum $SU(3)$ QCD, it is necessary to study  monopoles numerically in the framework of lattice $SU(3)$ QCD and to study then if the continuum limit exists. It is not so straightforward, however, to extend the previous $SU(2)$ studies to $SU(3)$. How to define Abelian-like link fields and monopoles without gauge-fixing is not so simple as  in $SU(2)$, since a $SU(3)$ group link field is not  expanded in terms of Lie-algebra elements defining Abelian link fields as simply done as in  $SU(2)$. There are theoretically many possible definitions which have the same naive continuum limit in $SU(3)$.  In this work, we introduce a natural definition of  the new-type of lattice Abelian-like fields and monopoles in $SU(3)$. 

The present paper is organized as follows. 
In Sec.\ref{theoretical background}, we first review shortly the theoretical background of our new Abelian-like monopoles published in Ref.\cite{Suzuki:2014wya}.  
Sec.\ref{lattice definition} is devoted to lattice descriptions of the definition of Abelian link field and Abelian-like monopole in $SU(3)$ QCD. In Sec.\ref{Abelian dominance}, \ref{monopole dominance},\ref{dual Meissner}, the results of numerical simulations on the lattice are shown. Our conclusions are given in Sec.\ref{conclusion}. 
In Appendix A, the problem how to define Abelian link fields out of non-Abelian one is discussed shortly.

\section{Equivalence of VNABI and Abelian-like monopoles} \label{theoretical background}
First of all, we shortly review the work~\cite{Suzuki:2014wya} that the Jacobi identities of covariant derivatives lead us to conclusion that violation of the non-Abelian Bianchi identities (VNABI) $J_{\mu}$ is nothing but an Abelian-like monopoles $k_{\mu}$ defined by violation of the Abelian-like Bianchi identities without gauge-fixing. 
Define a covariant derivative operator $D_{\mu}=\partial_{\mu}-igA_{\mu}$. The Jacobi identities are expressed as $\epsilon_{\mu\nu\rho\sigma}[D_{\nu},[D_{\rho},D_{\sigma}]]=0$.
By direct calculations, one gets
\begin{eqnarray*}
[D_{\rho},D_{\sigma}]&=&[\partial_{\rho}-igA_{\rho},\partial_{\sigma}-igA_{\sigma}]\\
&=&-igG_{\rho\sigma}+[\partial_{\rho},\partial_{\sigma}],
\end{eqnarray*}
where the second commutator term of the partial derivative operators can not be discarded in general, since gauge fields may contain a line singularity. Actually, it is the origin of the violation of the non-Abelian Bianchi identities (VNABI) as shown in the following. The non-Abelian Bianchi identities and the Abelian-like Bianchi identities are, respectively: $D_{\nu}G^{*}_{\mu\nu}=0$ and $\partial_{\nu}f^{*}_{\mu\nu}=0$.
The relation $[D_{\nu},G_{\rho\sigma}]=D_{\nu}G_{\rho\sigma}$ and the Jacobi identities lead us to
\begin{eqnarray}
D_{\nu}G^{*}_{\mu\nu}&=&\frac{1}{2}\epsilon_{\mu\nu\rho\sigma}D_{\nu}G_{\rho\sigma} \nn\\
&=&-\frac{i}{2g}\epsilon_{\mu\nu\rho\sigma}[D_{\nu},[\partial_{\rho},\partial_{\sigma}]]\nn\\
&=&\frac{1}{2}\epsilon_{\mu\nu\rho\sigma}[\partial_{\rho},\partial_{\sigma}]A_{\nu}
=\partial_{\nu}f^{*}_{\mu\nu}, \label{eq-JK}
\end{eqnarray}
where $f_{\mu\nu}$ is defined as $f_{\mu\nu}=\partial_{\mu}A_{\nu}-\partial_{\nu}A_{\mu}=(\partial_{\mu}A^a_{\nu}-\partial_{\nu}A^a_{\mu})\lambda^a/2$. Namely Eq.(\ref{eq-JK}) shows that the violation of the non-Abelian Bianchi identities, if exists,  is equivalent to that of the Abelian-like Bianchi identities.

Let us denote the violation of the non-Abelian Bianchi identities (VNABI) as  $J_{\mu}=D_{\nu}G^*_{\mu \nu}$ and Abelian-like monopole currents $k_{\mu}$ without any gauge-fixing as the violation of the Abelian-like Bianchi identities:
\begin{eqnarray}
k_{\mu}=\partial_{\nu}f^*_{\mu\nu}
=\frac{1}{2}\epsilon_{\mu\nu\rho\sigma}\partial_{\nu}f_{\rho\sigma}. \label{ab-mon}
\end{eqnarray}
Eq.(\ref{eq-JK}) shows that 
\begin{eqnarray}
J_{\mu}=k_{\mu}. \label{J-K}
\end{eqnarray}

Due to the antisymmetric property of the Abelian-like field strength, we get Abelian-like conservation conditions~\cite{Arafune:1974uy}:
\begin{eqnarray}
\partial_\mu k_\mu=0. \label{A-cons}
\end{eqnarray}

If such singularities exist actually in the continuum QCD, the Abelian dual Meissner effect could be the color confinement mechanism naturally. It is then very important to study the Abelian-like monopoles in the framework of lattice QCD.

As discussed in Introduction, the above authors' standpoint seems to work well at least in the framework of  $SU(2)$ QCD.  Abelian-like monopoles following DeGrand-Toussaint~\cite{DeGrand:1980eq} without additional gauge-fixing could reproduce the non-Abelian string tension perfectly for various coupling constants $\beta$ and lattice volumes as shown in Ref.\cite{Suzuki:2007jp,Suzuki:2009xy}. To
study the continuum limit more rigorously, the block-spin renormalization studies with respect to monopole operators after various smooth gauge-fixings could prove the existence of the gauge-invariant continuum limit of such Abelian-like monopoles~\cite{Suzuki:2017lco,Suzuki:2017zdh}. 

\section{Lattice study of $SU(3)$ QCD} \label{lattice definition}
First of all, we define Abelian link fields and  Abelian Dirac-type monopoles on $SU(3)$ lattice.

\subsection{Defining an Abelian link field $\theta_\mu^a$ from non-Abelian link field $U_\mu(s)$}
In the usual lattice $SU(3)$ QCD, a non-Abelian link field $U_\mu(s)$ as a $SU(3)$ group element is used as a dynamical quantity defined on a link $(s,\mu)$. How to extract an Abelian link field $\theta_\mu^a$ having a color $a$ out of $U_\mu(s)$ is not trivial especially without any additional partial gauge-fixing to Abelian torus group $U(1)\times U(1)$. There are many possible ways leading naively to the same $a\to0$  continuum limit. We find that the following simple method is a good candidate being consistent with the previous $SU(2)$ method adopted in Ref.\cite{Suzuki:2009xy,Suzuki:2017lco,Suzuki:2017zdh}. 
The situations behind the choice are discussed in Appendix A. 
As discussed in Appendix A, we fix them to maximize the following quantity locally
\begin{eqnarray}
RA= \mathrm{Re} \tr\left\{\exp(i\theta_\mu^a(s)\lambda^a)U_\mu^{\dag}(s)\right\}, \label{RA}
\end{eqnarray}
where $\lambda^a$ is the Gell-Mann matrix and no sum over $a$ is not taken.

\subsubsection{The $SU(2)$ case}
In $SU(2)$, Eq.(\ref{RA}) leads us to
\begin{eqnarray}
\theta_\mu^a(s) = \tan^{-1}\left\{\frac{U_\mu^a(s)}{U_\mu^0(s)}\right\},\label{theta2}
\end{eqnarray}
where $U_\mu(s)=U_\mu^0(s)+ i\sum_a\sigma^aU_\mu^a(s)$. This definition is the same as that used in the previous works~\cite{Suzuki:2007jp,Suzuki:2009xy} where Abelian and monopole dominances were proved numerically without adopting new additional  gauge-fixings.

\subsubsection{The $SU(3)$ case}
Eq.(\ref{RA}) gives us in $SU(3)$ the following Abelian link fields for example in the $\lambda^1$ case: 
\begin{eqnarray}
\theta^1_\mu(s)=\tan^{-1}\left\{\frac{\mathrm{Im}(U_{12}(s,\mu)+U_{21}(s,\mu))}{\mathrm{Re}(U_{11}(s,\mu)+U_{22}(s,\mu))}\right\}.  \label{Eq1}
\end{eqnarray}
Other  $a=2\sim 7$ cases are fixed similarly. But in the case of $\lambda^8$, 
we have to maximize 
\begin{eqnarray}
R^8&=&\cos t_1 \mathrm{Re}(U_{11}+U_{22})+\cos 2t_1 \mathrm{Re}(U_{33}) \nonumber \\
 &+&\sin t_1 \mathrm{Im}(U_{11}+U_{22})-\sin 2t_1 \mathrm{Im}(U_{33}),  \label{Eq81} 
\end{eqnarray}
where $t_1=\theta^8/2$. The maximization of Eq.(\ref{Eq81}) gives us a quartic equation with respect to $t_2=\tan(\theta^8/(2\sqrt{3}))$. The quartic equation is easy to solve rather numerically but to make the solution compact between $\left[-\pi,\pi\right]$, we redefine in the following way:
\begin{eqnarray}
\theta^8=\tan^{-1}\frac{2t_2}{1-t_2^2}, \label{Eq8}
\end{eqnarray}
where the range is extended to $\left[-\pi,\pi\right]$.  With respect to two diagonal parts, there are other Weyl symmetric definition which gives us different numerical results for finite $a(\beta)$. But in this paper, we adopt the above definition (\ref{Eq8}) for simplicity, since numerically no big difference is found. 

\subsection{Definition of Abelian lattice monopoles}
Now that the Abelian link fields are defined, we next define Abelian lattice monopoles. The unique reliable method ever known to define a lattice Abelian monopole is the one proposed in compact QED by DeGrand and Toussaint~\cite{DeGrand:1980eq} who utilize the fact that the Dirac monopole has a Dirac string with a magnetic flux satisfying the Dirac quantization condition. Hence we adopt the method here, since the Abelian-like monopoles here are of the Dirac type in QCD. 

First we define Abelian plaquette variables from the above Abelian link variables:
\begin{eqnarray}
\theta_{\mu\nu}^a(s)&\equiv&\partial_{\mu}\theta_{\nu}^a(s)-\partial_{\nu}\theta_{\mu}^a(s),
\label{abel_proj}
\end{eqnarray}
where $\partial_{\nu}(\partial'_{\nu})$ is a forward (backward) difference. Then the plaquette variable can be decomposed as follows:
\begin{eqnarray}
\theta_{\mu\nu}^a(s) &=&\bar{\theta}_{\mu\nu}^a(s)+2\pi
n_{\mu\nu}^a(s)\ \ (|\bar{\theta}_{\mu\nu}^a|<\pi),\label{abel+proj}
\end{eqnarray}
where $n_{\mu\nu}^a(s)$ is an integer
corresponding to the number of the Dirac string.
Then VNABI as Abelian monopoles is defined by
\begin{eqnarray}
k_{\mu}^a(s)&=& -\frac{1}{2}\epsilon_{\mu\alpha\beta\gamma}\partial_{\alpha}
\bar{\theta}_{\beta\gamma}^a(s+\hat\mu) \nonumber\\
&=&\frac{1}{2}\epsilon_{\mu\alpha\beta\gamma}\partial_{\alpha}
n_{\beta\gamma}^a(s+\hat\mu), \nonumber \\
J_{\mu}(s)&\equiv&\frac{1}{2}k_{\mu}^a(s)\lambda^a \label{eq:amon}.
\end{eqnarray}
This definition (\ref{eq:amon}) of VNABI satisfies the Abelian conservation condition (\ref{A-cons}) and takes an integer value
which corresponds to the magnetic charge obeying the Dirac quantization
condition.  

\section{Perfect Abelian dominance} \label{Abelian dominance}
First of all, we calculate Abelian static potentials using the Abelian link variables (\ref{Eq1}). We generate thermalized gauge configurations using the $SU(3)$ Wilson action 
at coupling constants~$\beta=5.6, 5.7$ and $5.8$ where the lattice spacings ~$a(\beta=5.6)=0.2235$~[fm], ~$a(\beta=5.7)=0.17016$~[fm] and ~$a(\beta=5.8)=0.13642$~[fm]
are cited from Ref.\cite{NS:2001}. The lattice sizes are $N_{s}^3 \times N_{t}=12^3 \times 12$ at $\beta=5.6, 5.7, 5.8$ and $16^3 \times 16$ at $\beta=5.6$. 
\par
By using the multi-level noise reduction 
method~\cite{Luscher:2001up, Luscher2002, Koma2006, Koma2007, Koma2017}, we
evaluate the Abelian static potential $V_{\rm A}$ from the 
correlation function (PLCF) of the 
Abelian Polyakov loop operator
\begin{equation}
P^a_{\rm A} = \exp[i\sum_{k=0}^{N_{t}-1}\theta^a_4(s+k\hat{4})] \;,
\label{eq-PA}
\end{equation}
separated at a distance $r$ as
\begin{equation}
V^a_{\rm A}(r) = -\frac{1}{a N_{t}}\ln \langle  P^a_{A}(0) P_{A}^{*a}(r)\rangle \;.
\end{equation}
The sublattice sizes adopted are $2a$ at $\beta=5.6, 5.7$ and $3a$ at $\beta=5.8$.
The parameters for the multilevel algorithm here we used are summarized in Table~\ref{SU(3)multilevel_parameter}.

\begin{figure}[htbp]
  \begin{center}
  \hspace*{1.5cm} 
  \includegraphics[width=7cm]{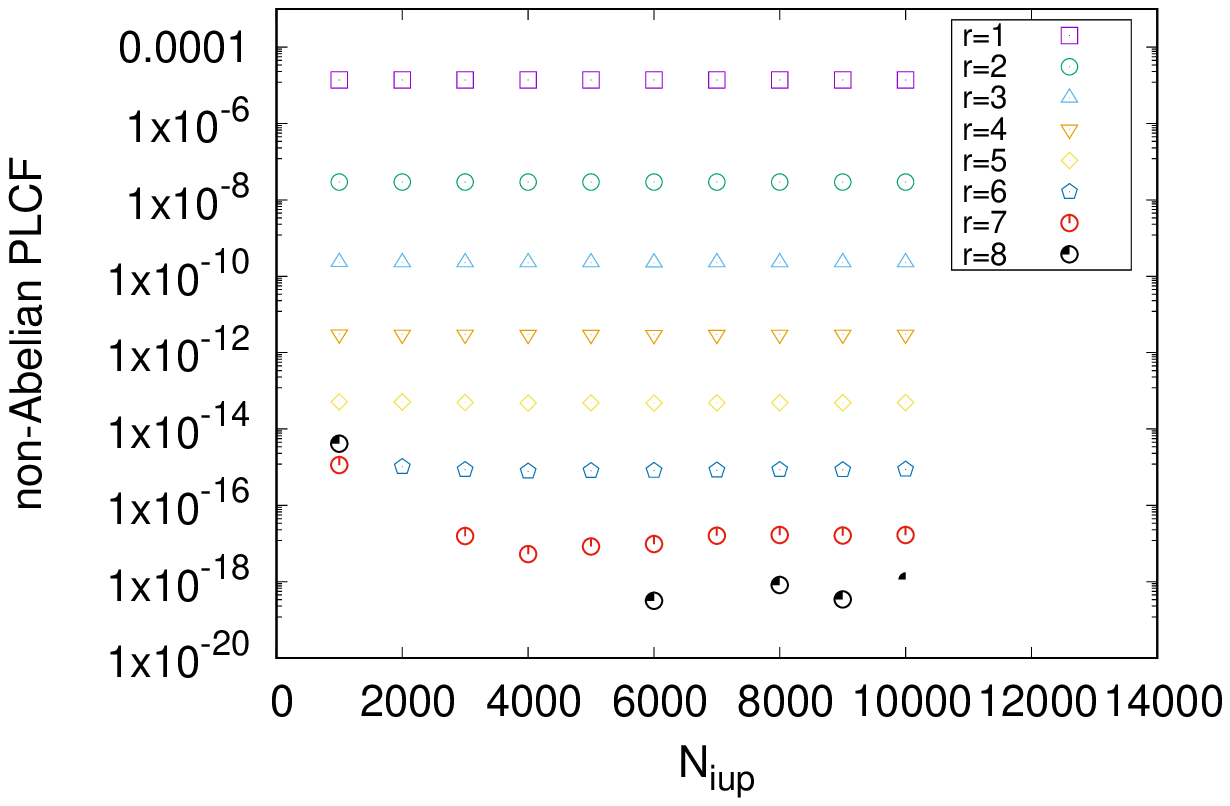}
  \end{center}
  \vspace*{.5cm} 
  \caption{ Convergence history of the non-Abelian PLCF at $\beta=5.6$ on $16^3\times 16$ lattice. }\label{Convergence_16xx4_b560_f}
\end{figure}

\begin{figure}[htbp]
  \begin{center}
  \hspace*{1.5cm} 
  \includegraphics[width=7cm]{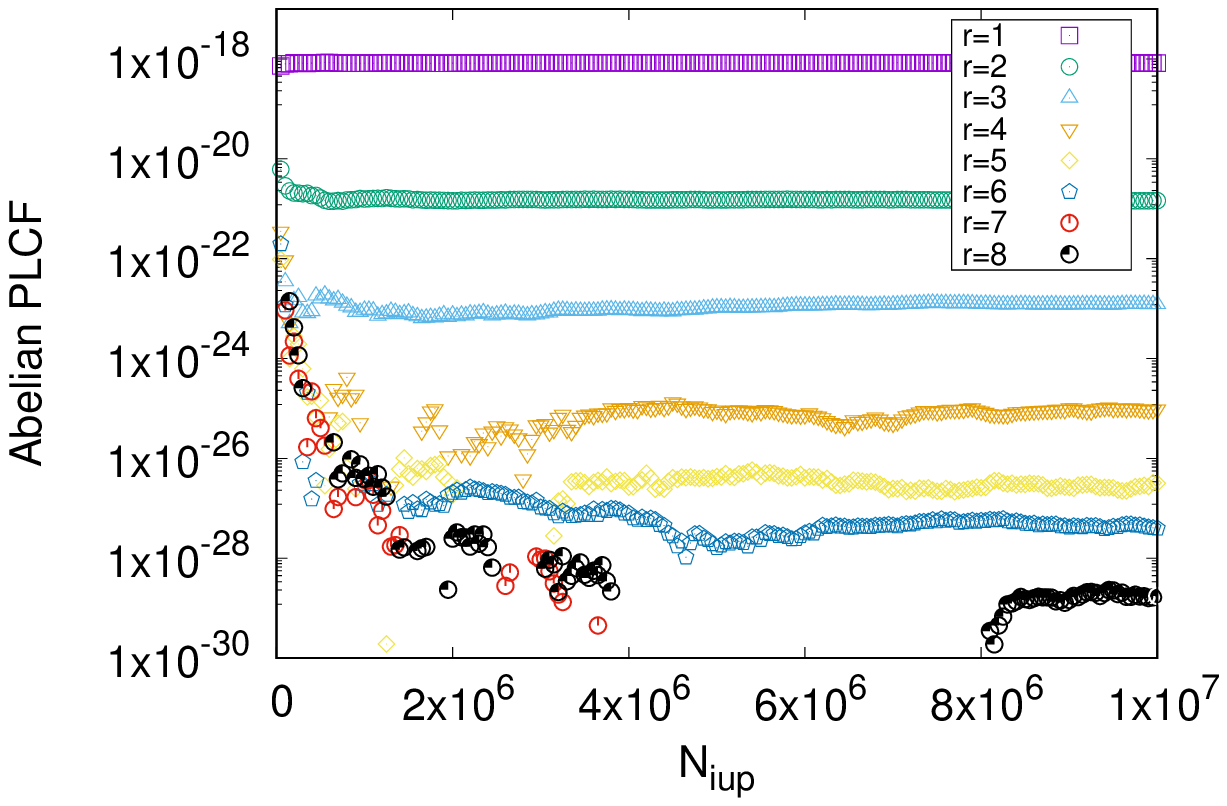}
  \end{center}
  \vspace*{.5cm} 
  \caption{ Convergence history of the Abelian PLCF at $\beta=5.6$ on $16^3\times 16$ lattice. }\label{Convergence_16xx4_b560_a}
\end{figure}

\begin{figure}[htbp]
\begin{center}
\hspace*{1.5cm} 
\includegraphics[width=7cm]{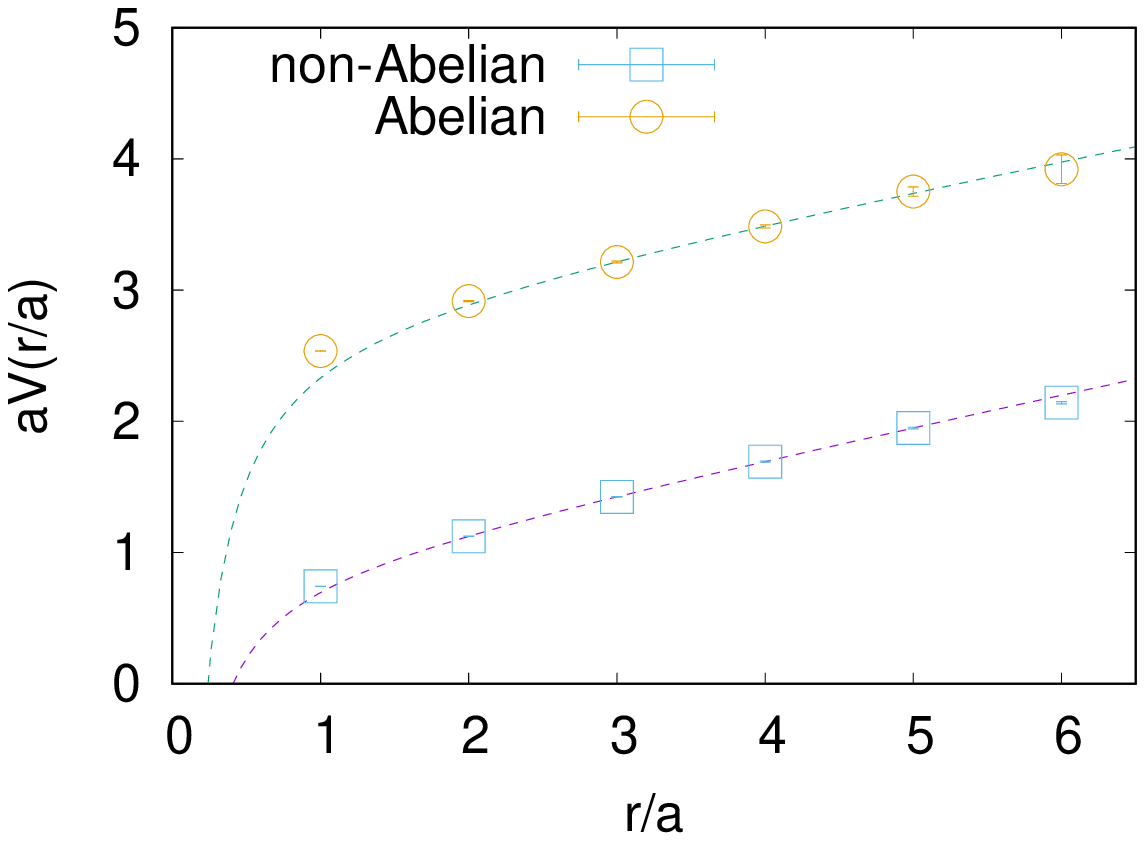}
\end{center}
\vspace*{.5cm} 
\caption{The static quark potentials from non-Abelian and Abelian PLCF at $\beta = 5.6$ on $12^3\times 12$ lattice.}\label{POTENTIAL_12xx4_b560}
\end{figure}

\begin{figure}[htbp]
\begin{center}
\hspace*{1.5cm} 
\includegraphics[width=7cm]{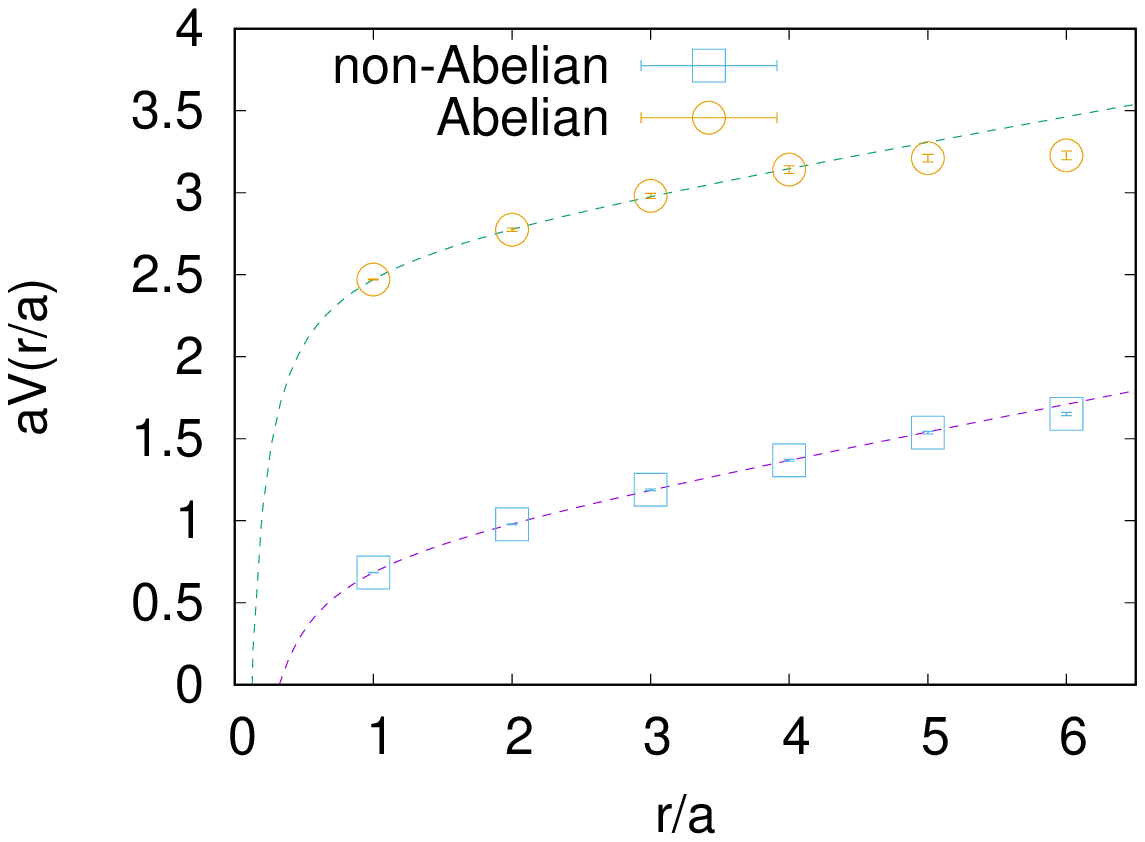}
\end{center}
\vspace*{.5cm} 
\caption{The static quark potentials from non-Abelian and Abelian PLCF at $\beta = 5.7$ on $12^3\times 12$ lattice.}\label{POTENTIAL_12xx4_b570}
\end{figure}

\begin{figure}[htbp]
\begin{center}
\hspace*{1.5cm} 
\includegraphics[width=7cm]{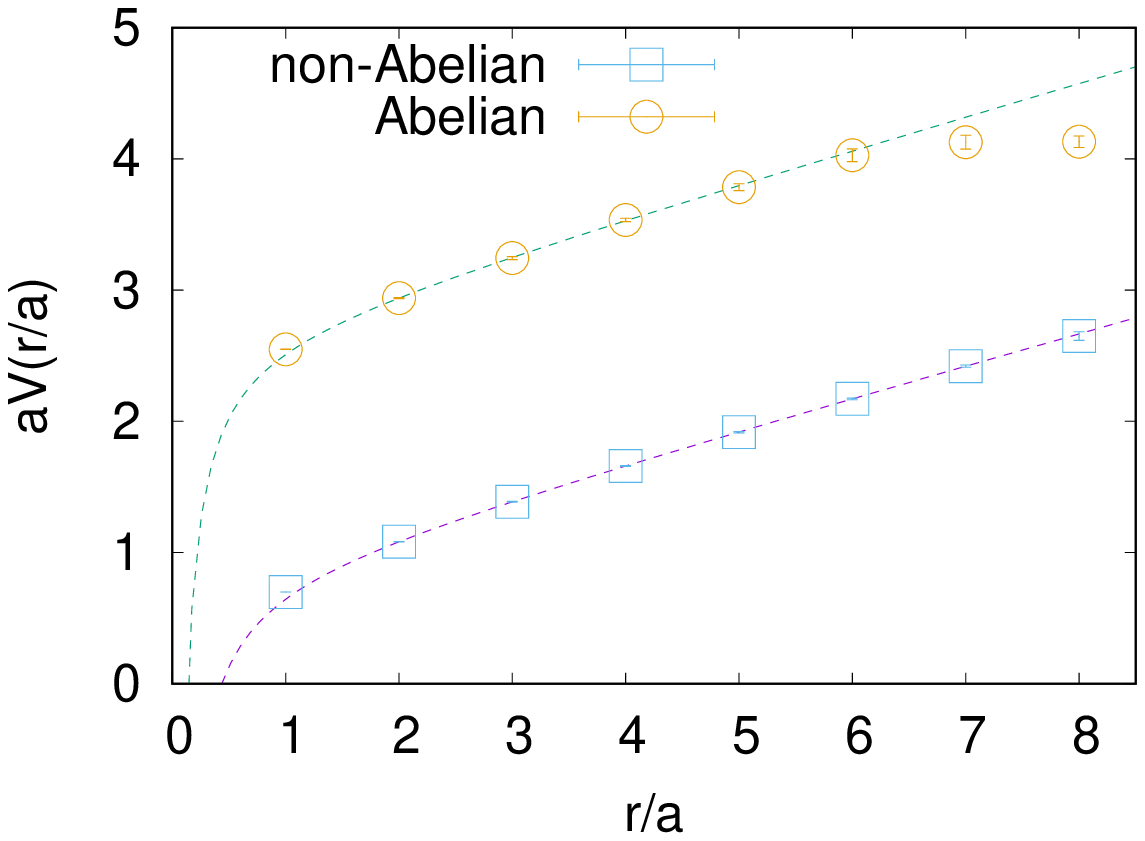}
\end{center}
\vspace*{.5cm} 
\caption{The static quark potentials from non-Abelian and Abelian PLCF at $\beta = 5.6$ on $16^3\times 16$ lattice.}\label{POTENTIAL_16xx4_b560}
\end{figure}

We show convergence behaviors in a configuration with respect to internal updates $N_{\rm iup}$ concerning a non-Abelian PLCF in Fig.~\ref{Convergence_16xx4_b560_f} and also an Abelian PLCF in Fig.~\ref{Convergence_16xx4_b560_a} on  $16^3\times 16$ lattice at $\beta=5.6$.
From Fig.~\ref{Convergence_16xx4_b560_f}, we get almost convergence around 
$N_{\rm iup}\sim 10^4$ in non-Abelian PLCF. On the other hand, in the case of Abelian PLCF, even around $N_{\rm iup}\sim 10^7$, convergence is not good enough for large $r>6$.  Due to the limited computer resources,  we fix $N_{\rm iup}=10^7$ and try to increase number of configurations as much as possible. Since global color symmetry is not broken, we adopt only $a=1$ color case in this calculation.
The results are fairly good as seen from Fig.~\ref{POTENTIAL_12xx4_b560}$\sim$\ref{POTENTIAL_16xx4_b560}.  
We see a flattening behavior at $r\ge 5$ in Fig.~\ref{POTENTIAL_12xx4_b570} and at $r\ge 7$ in 
Fig.~\ref{POTENTIAL_16xx4_b560} respectively due to insufficient number of internal updates $N_{\rm iup}$. Since increasing the number of internal updates more is impossible, we tune the fitting range at smaller $r$ region as shown in Table~\ref{stringtension_multilevel}. 
We try to fit the data to a usual function $V(r) = \sigma r- c/r+\mu$ and find almost the same string tension $\sigma$ and the Coulombic coefficient $c$ as shown in Table~\ref{stringtension_multilevel}, indicating almost perfect Abelian dominance. 
Here the number of independent vacuum configurations is $6$ in all cases. The errors are determined by the jackknife method. Results similar to those obtained in the case of $SU(2)$ gauge theory~\cite{Suzuki:2007jp, Suzuki:2009xy} are shown also in the case of $SU(3)$ gauge theory. Our results of the string tension are consistent with theoretical observations on the basis of reasonable assumptions~\cite{Ogilvie:1998wu,Faber:1998en}. 

\begin{table}[t]
\begin{center}
\caption{\label{SU(3)multilevel_parameter}
Simulation parameters for the measurement of static potential using the multilevel method.
$N_{\rm sub}$ is the sublattice size divided and $N_{\rm iup}$ is the number of internal updates in the multilevel method.}
\begin{tabular}{c|c|c|c|c|c}
\hline
$\beta$ &$N_{s}^{3}\times N_{t}$& $a(\beta)$~[fm]& $N_{\rm conf}$ & $N_{\rm sub}$ & $N_{\rm iup}$ \\ 
\hline
5.60 & $12^3 \times 12$ & 0.2235\F & 6 & 2 & 5000000\\
5.60 & $16^3 \times 16$ & 0.2235\F & 6 & 2 & 10000000\\
5.70 & $12^3 \times 12$ & 0.17016\F & 6 & 2 & 5000000\\
5.80 & $12^3 \times 12$ & 0.13642\F & 6 & 3 & 5000000\\
\hline
\end{tabular}
\end{center}
\end{table}

\begin{figure}[htbp]
\begin{center}
\hspace*{1.5cm} 
\includegraphics[width=7cm]{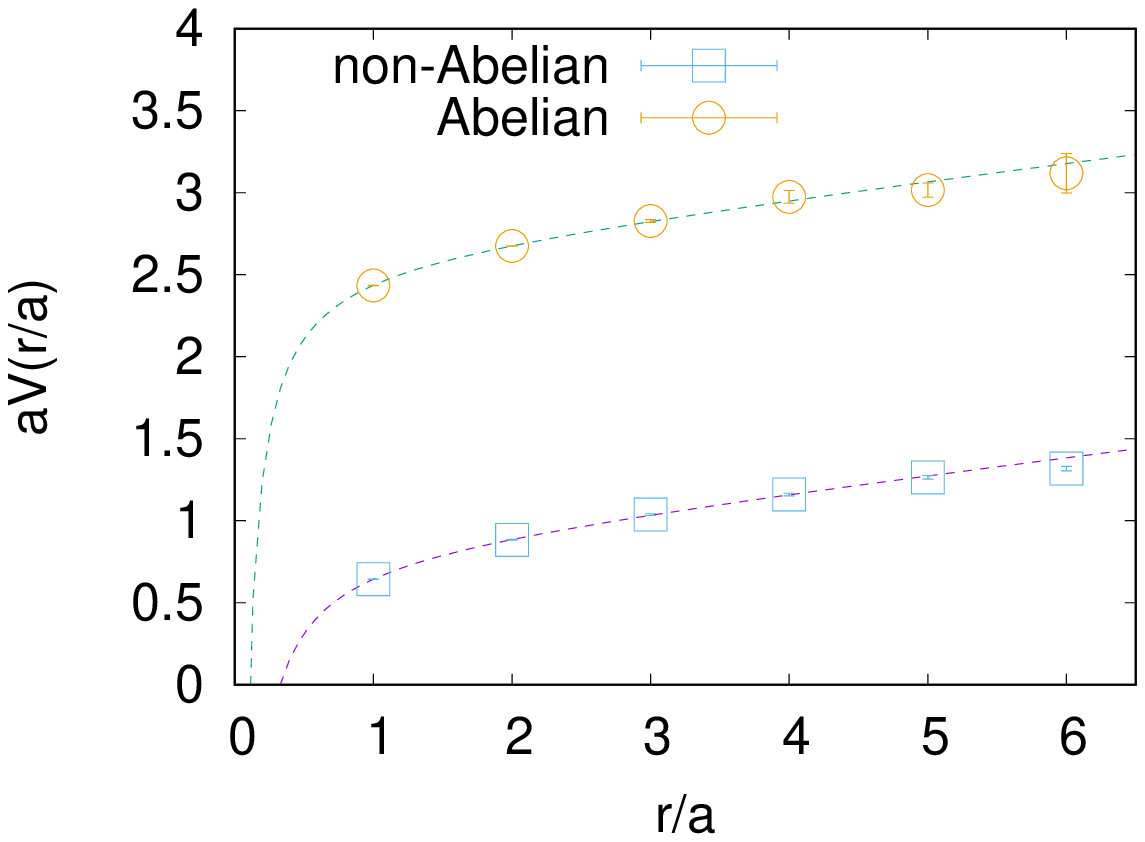}
\end{center}
\vspace*{.5cm} 
\caption{The static quark potentials from non-Abelian and Abelian PLCF at $\beta = 5.8$ on $12^3\times 12$ lattice.}\label{POTENTIAL_12xx4_b580}
\end{figure}

\begin{table}[t]
\begin{center}
\caption{
Best fitted values of the string tension $\sigma a^2$, the
Coulombic coefficient $c$, and the constant $\mu a$ for the
potentials $V_{\rm NA}$, $V_{\rm A}$. FR means the fitting range.}
\label{stringtension_multilevel}
\begin{tabular}{l|c|c|c|c|c}
\hline
& $\sigma a^2$ & $c$ & $\mu a$ & FR($r/a$) & $\chi^2/N_{\rm df}$ \\ 
\hline
\multicolumn{6}{l}{$\beta =5.6, 12^3\times 12$}\\
\hline
$V_{\rm NA}$   & 0.2368(1)  & -0.384(1)  & 0.8415(7)  & 2 - 5 & 0.0004  \\ 
$V_{\rm A}$    & 0.21(5)    & -0.6(6)    & 2.7(4)     & 3 - 6 & 0.42    \\ 
\hline
\multicolumn{6}{l}{$\beta =5.6, 16^3\times 16$}\\ 
\hline
$V_{\rm NA}$  & 0.239(2)  & -0.39(4)  & 0.79(2)  & 3 - 8 & 0.0903 \\ 
$V_{\rm A}$   & 0.25(2)   &  -0.3(1)  & 2.6(1)   & 2 - 5 & 0.6044 \\
\hline
\multicolumn{6}{l}{$\beta =5.7, 12^3\times 12$}\\\hline
$V_{\rm NA}$  & 0.159(3)  & -0.272(8)  & 0.79(1)  & 1 - 5 & 0.5362 \\ 
$V_{\rm A}$   & 0.145(9)  & -0.32(2)   & 2.64(3)  & 1 - 4 & 0.2226 \\ 
\hline
\multicolumn{6}{l}{$\beta =5.8, 12^3\times 12$}\\ 
\hline
$V_{\rm NA}$  & 0.101(3)  & -0.28(1)  &  0.82(1)  & 1 - 5 & 0.8013 \\ 
$V_{\rm A}$   & 0.102(9)  & -0.27(2)  &  2.60(3)  & 1 - 5 & 0.9993 \\
\hline
\end{tabular}
\end{center}
\end{table}
Due to insufficient computer resources, the continuum limit and the
infinite volume limit are not studied yet, and estimates of various systematic errors are incomplete. Nevertheless, the results obtained are very interesting because they are the first to show Abelian dominance for the string tension in $SU(3)$ gauge theory without any additional gauge fixing.

\section{Perfect monopole dominance} \label{monopole dominance}
If the Abelian dual Meissner effect due to Abelian monopole currents is 
the essence of color confinement in QCD, the $SU(3)$ string tension is reproduced 
completely by a solenoidal current due to the Abelian monopole. Namely the so-called perfect monopole dominance is expected to occur with respect to the string tension.

Already in $SU(3)$ QCD, almost perfect monopole dominance is shown in MA gauge in a restricted case~\cite{Suganuma:2018}.  Here we investigate the monopole contribution without any gauge-fixing to the static potential in order to examine the role of monopoles for confinement in a gauge independent way.
The monopole part of the Polyakov loop operator is extracted as follows.
Using the lattice Coulomb propagator $D(s-s')$, which satisfies
$\partial_{\nu}\partial'_{\nu}D(s-s') = -\delta_{ss'}$, 
the temporal component of the Abelian fields $\theta^a_{4}(s)$ are written as 
\begin{equation}
\theta^a_4 (s) 
= -\sum_{s'} D(s-s')[\partial'_{\nu}\theta^a_{\nu 4}(s')+
\partial_4 (\partial'_{\nu}\theta^a_{\nu}(s'))] \; . 
\label{t4}
\end{equation}
Inserting Eq.\eqref{t4} and then Eq.\eqref{abel+proj}
to the Abelian Polyakov loop~\eqref{eq-PA},
we obtain
\begin{eqnarray}
&&P^a_{\rm A} = P^a_{\rm ph} \cdot P^a_{\rm mon}\; ,\nonumber\\
&&P^a_{\rm ph} = \exp\{-i\sum_{k=0}^{N_{t}-1} \!\sum_{s'}
D(s+k\hat4-s')\partial'_{\nu}\bar{\theta}^a_{\nu 4}(s')\} \; ,\nonumber\\
&&P^a_{\rm mon} = \exp\{-2\pi i\sum_{k=0}^{N_{t}-1}\! \sum_{s'}
D(s+k\hat4-s')\partial'_{\nu}n^a_{\nu 4}(s')\}\; .\nonumber\\
\label{ph-mon}
\end{eqnarray}
We call $P^a_{\rm ph}$ the photon
and $P^a_{\rm mon}$ the monopole parts of 
the Abelian Polyakov loop $P^a$, respectively~\cite{Suzuki:1994ay}.
The latter is due to the fact that the Dirac strings 
$n^a_{\nu 4}(s)$ lead to the monopole currents in 
Eq.\eqref{eq:amon}~\cite{DeGrand:1980eq}.
Note that the second term of Eq.~\eqref{t4} does
not contribute to the Abelian Polyakov loop 
in Eq.\eqref{eq-PA}.

Since Eq.\eqref{ph-mon} contains
the non-local Coulomb propagator $D(s-s')$
and the Polyakov loop is not written as 
a product of local operators along the time direction,
the above multilevel method~\cite{Luscher:2001up} cannot be applied.
Without such a powerful noise reduction method, 
it is hard to measure the Polyakov loop correlation function
at zero temperature with the present available computer resource.
Thus we consider a finite temperature $T \neq 0$ system 
in the confinement phase.
We set $T = 0.8T_{c}$ and simulate the Wilson action on the $24^3 \times 4$ lattice with $\beta=5.6$.  To check scaling, we tried to do simulations on lattices having the time distances $N_t=6$ ($\beta=5.75$) and $N_t=8$ ($\beta=5.9$) also corresponding to the same $T = 0.8T_{c}$.  Unfortunately however it is found that we need too large number of vacuum configurations to get meaningful results on such larger lattices. Hence we restrict ourselves to the above smallest lattice.  

\subsection{Noise reduction by gauge averaging and simulation parameters}
Since the signal-to-noise ratio of the correlation functions of
$P_{\rm A}$, $P_{\rm ph}$ and $P_{\rm mon}$ are very small without  any smooth gauge fixing, we adopt a  noise reduction method~\cite{Suzuki:2007jp}.
For a thermalized gauge configuration,
we produce many gauge copies applying random gauge transformations.
Then we compute the operator for each copy, and 
take the average over all copies.
It should be noted that as long as a gauge-invariant operator is evaluated, 
such copies are identical, but they are not if a gauge-variant 
operator is evaluated as in the present case. Also since the global color invariance  exists with respect to  colors of Abelian monopoles, we include the different color contributions into the average.    The results obtained with this method are gauge-averaged and so  gauge-invariant. We show the simulation parameters of the $SU(3)$ case and for comparison, as well as previous  $SU(2)$ case~\cite{Suzuki:2009xy} in Table~\ref{SU2-SU3data}. 
In practice, we prepare a few hundred or a thousand of gauge 
copies for each independent gauge configuration (see Table~\ref{SU2-SU3data}).
We also apply one-step hypercubic
blocking (HYP)~\cite{Hasenfratz:2001hp}
to the temporal links for further noise reduction.
The short-distance part of the potential may be affected by HYP.

\begin{table}
\begin{center}
\caption{\label{SU2-SU3data}
Simulation parameters for the measurement of the static potential 
from $P_{\rm A}$, $P_{\rm ph}$ and $P_{\rm mon}$ in $SU(3)$ and $SU(3)$.
$N_{\rm RGT}$ is the number of random gauge transformations. The $SU(2)$ data are cited from Ref.\cite{Suzuki:2009xy}.}
\begin{tabular}{c|c|c|c|c|c}
\hline
&$\beta$ &$N_{s}^{3}\times N_{t}$& $a(\beta)$~[fm]& $N_{\rm conf}$ & $N_{\rm RGT}$ \\ 
\hline
$SU(3)$&5.60 & $24^{3} \times 4$ & 0.2235\F & 910000 & 400\\
\hline
$SU(2)$&2.43 &$24^{3} \times 8$ & 0.1029(4)& 7000 & 4000 \\
\hline
\end{tabular}
\end{center}
\end{table}

\begin{table}
\centering
\fontsize{8pt}{20pt}\selectfont
\caption{Best fitted values of the string tension $\sigma a^2$, the
Coulombic coefficient $c$, and the constant $\mu a$ for the
potentials $V_{\rm NA}$, $V_{\rm A}$, $V_{\rm mon}$ and $V_{\rm ph}$. Here $V_{\rm mon}$ 
in $SU(3)$ alone is fitted in terms of $V(r)=\sigma r + \mu$. Others are fitted by $V(r)=\sigma r  -c/r +\mu$.  FR means the fitting range.
One of the $SU(2)$ data are cited for comparison from Ref.\cite{Suzuki:2009xy}.}
\label{FAMPFIT}
\begin{tabular}{l|c|c|c|c|c}
\hline
\multicolumn{6}{l}{$SU(3)$ ($24^3\times 4$)}\\ 
\hline
& $\sigma a^2$ & $c$ & $\mu a$ & FR($r/a$) 
& $\chi^2/N_{\rm df}$ \\ \hline
$V_{\rm NA}$   & 0.178(1)    & 0.86(4)  & 0.99(1)  & 5 - 9  & 1.23 \\ 
$V_{\rm A}$    & 0.16(3)     & 0.9(11)  & 2.5(3)   & 5 - 9  & 1.03 \\ 
$V_{\rm mon}$  & 0.17(2)     &          & 2.9(1)   & 4 - 7  & 1.08 \\ 
$V_{\rm ph}$   &$-0.0007(1)$ & 0.046(3) & 0.945(1) & 3 - 10 & 7.22e-08 \\
\hline
\multicolumn{6}{l}{$SU(2)$ ($24^3\times 8$)}\\ \hline
$V_{\rm NA}$  & 0.0415(9)              & 0.47(2)  & 0.46(8)    & 4.1 - 7.8\F & 0.99 \\ 
$V_{\rm A}$   & 0.041(2)               & 0.47(6)  & 1.10(3)    & 4.5 - 8.5\F & 1.00 \\ 
$V_{\rm mon}$ & 0.043(3)               & 0.37(4)  & 1.39(2)    & 2.1 - 7.5\F & 0.99 \\ 
$V_{\rm ph}$  &$-6.0(3)\times 10^{-5}$ &0.0059(3) & 0.46649(6) & 7.7 - 11.5  & 1.02\\
 \hline
\end{tabular}
\end{table}

\subsection{Static potentials}
\begin{figure}[htbp]
\begin{center}
\hspace*{1.5cm} 
\vspace*{1cm} 
\includegraphics[width=7cm]{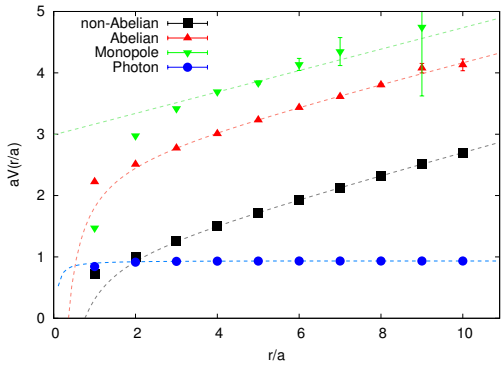}
\includegraphics[width=7.5cm]{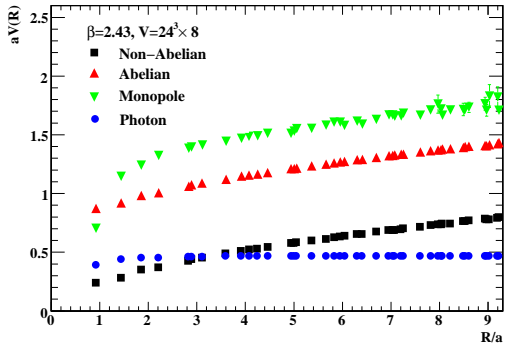}
\end{center}
\caption{The $SU(3)$ static quark potentials from PLCF at $\beta = 5.60$ on $24^3\times 4$ lattice (up). For comparison, the $SU(2)$ static quark potentials from PLCF at $\beta = 2.43$ on $24^3\times 8$ lattice (down) cited from Ref.\cite{Suzuki:2009xy}.} \label{POT24xx3x4}
\end{figure}

$SU(3)$ studies are found to be very much difficult and time consuming as seen from
Table~\ref{SU2-SU3data}.  We need much more gauge configurations than expected from the previous $SU(2)$ study before getting meaningful signal to noise ratio. In the case of 
$N_s^3\times N_t=24^3\times 4$, we need to use about a million configurations with additional $400$ random gauge copies per each. 
Since global color invariance is not broken, we take averages over  $a=1\sim 5$ five colors to improve statistics.
We get data suggesting perfect Abelian and monopole dominances 
as shown in Table~\ref{FAMPFIT}.
We obtain good signals for the Abelian, the monopole and the photon contributions to the static potential as shown in Fig.\ref{POT24xx3x4}. We try to fit the potentials in Fig.\ref{POT24xx3x4} to the function $V(r)=\sigma r -c/r+\mu$ and extract the string tension and the Coulombic coefficient of each potential as summarized in Table \ref{FAMPFIT}. Here $V_{\rm mon}$ alone is fitted in terms of $V(r)=\sigma r +\mu$. Abelian dominance is seen again in this case. Moreover, we can see monopole dominance, namely, only the monopole part of PLCF is responsible for the string tension. The photon part has no linear potential. 

\begin{table}[htbp]
\begin{center}
\caption{\label{E-k2}
Simulation parameters for the measurement of Abelian color electric fields $E^a_i$ and monopole currents $k^2$. 
$N_{\rm conf}$, $N_{\rm RGT}$ and $N_{\rm sm}$ are numbers of configurations, random gauge transformations and smearing, respectively. }
\begin{tabular}{c|c|c|c}
\hline
\multicolumn{4}{l}{$E^a_i$}\\
\hline
    &$N_{\rm conf}$& $N_{\rm RGT}$ & $N_{\rm sm}$ \\ 
\hline
 d=3& 20000&100&90\\
 d=4& 20000& 100& 90\\
 d=5& 80000& 100& 120\\
 d=6& 80000& 100& 120\\
\hline
\multicolumn{4}{l}{$k^2$}\\
\hline
    &$N_{\rm conf}$& $N_{\rm RGT}$ & $N_{\rm sm}$ \\ 
\hline
 d=3& 80000&0&90\\
 d=4& 160000& 0& 90\\
 d=5& 960000& 0& 120\\
 d=6& 960000& 0& 120\\
\hline
\end{tabular}
\end{center}
\end{table}

\begin{table}[htbp]
\begin{center}
\caption{\label{rotE-k}
Simulation parameters for the measurement of the dual Amp\`{e}re's law and Abelian monopole currents $k^a_i$ 
for the distance $d=3$. 
$N_{\rm conf}$, $N_{\rm RGT}$ and $N_{\rm sm}$ are numbers of configurations, random gauge copies and smearing, respectively. }
\begin{tabular}{c|c|c|c}
\hline
     &$N_{\rm conf}$& $N_{\rm RGT}$& $N_{\rm sm}$ \\ 
\hline
$(\mathrm{rot}E^{a})_\phi$ and $\partial_{t}B^{a}_{\phi}$ & 20000&100&90\\
\hline
$k^a_{\phi}$& 11200&3000&90\\
\hline
$k^a_r$ and  $k^a_z$ & 9600&3000&90\\
\hline 
\end{tabular}
\end{center}
\end{table}

\section{The Abelian dual Meissner effect in $SU(3)$} \label{dual Meissner}
\subsection{Simulation details of the flux-tube profile}
In this section, we show the results with respect to the Abelian dual Meissner effect. In the previous work~\cite{Suzuki:2009xy} studying the spatial distribution of color electric fields and monopole currents, they  used the connected correlations between a non-Abelian Wilson loop and Abelian operators in $SU(2)$ gauge theory without gauge fixing. We apply the same method to $SU(3)$ gauge theory without gauge fixing. Here we employ the standard Wilson action on the $24^3 \times 4$ lattice with the coupling constant $\beta = 5.60$ as done in the previous section.
To improve the signal-to-noise ratio, the APE smearing is applied to the spatial links and the hypercubic blocking~\cite{Hasenfratz:2001hp} 
is applied to the temporal links. We introduce random gauge transformations to improve the signal to noise ratios of the data concerning the Abelian operators. All simulation parameters are listed in Tables \ref{E-k2} and \ref{rotE-k}. 

\begin{figure}[h]
    \centering
\includegraphics[keepaspectratio, scale=0.7]{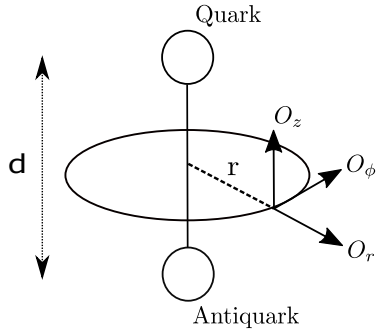}
 \vspace*{.5cm} 
\caption{The cylindrical coordinate}\label{cor}  
\end{figure}

To measure the flux-tube profiles, we consider a connected correlation functions as done in \cite{CCA:PRD89, Cea2016, refId0, BB:PRD99}:  
\begin{eqnarray}
\rho_{conn}(O(r))& = &\frac{\kbra{\Tr(P(0)LO(r)L^{\dagger})\Tr P^{\dagger}(d)}}{\kbra{\Tr P(0) \Tr P^{\dagger}(d)}} \nonumber\\
&&   - \frac{1}{3}\frac{\kbra{\Tr P(0) \Tr P^{\dagger}(d) \Tr O(r)}}{\kbra{\Tr P(0) \Tr P^{\dagger}(d)}}, \label{connect}
\end{eqnarray} 
where $P$ denotes a non-Abelian Polyakov loop, $L$ indicates a Schwinger line, $r$ is a distance from a flux-tube and $d$ is a distance between Polyakov loops. We use the cylindrical coordinate $(r,\phi,z)$ to parametrize the $q\text{-}\bar{q}$ system as shown in Fig.\ref{cor}. Here  the definition of the cylindrical coordinate $(r, \phi, z)$ along the $q\text{-}\bar{q}$ axis is shown.
 
\begin{figure}[h]
      \centering
      \hspace*{1cm} 
    \includegraphics[keepaspectratio, scale=0.6]{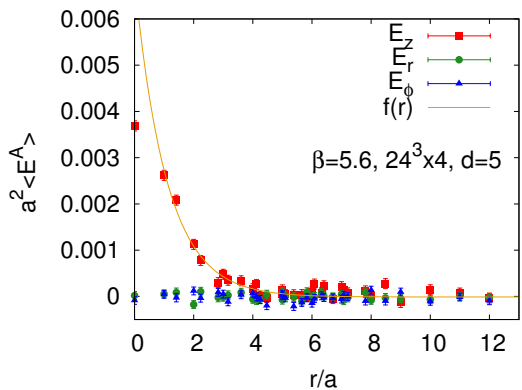} 
     \vspace*{.5cm} 
  \caption{The Abelian color electric field around static quarks for $d=5$ at $\beta = 5.6$ on $24^3\times 4$ lattices.}\label{EA_zrp} 
\end{figure}
\begin{table}[h] 
\begin{center}
\caption{The penetration length $\lambda$ of Abelian color electric fields at $\beta = 5.6$ on $24^3 \times 4$ lattices.} \label{lambda_A}
\begin{tabular}{|c|c|c|c|c|} \hline
 \multicolumn{1}{|c|}{$d$} &  \multicolumn{1}{|c|}{$\lambda/a$}&  \multicolumn{1}{|c|}{$c_1$}  &  \multicolumn{1}{|c|}{$c_0$}&$\chi^2/N_{\rm df}$ \\ \hline
 3&0.91(1)&0.0100(2)&-0.000002(8)&1.31628\\
\hline
4&1.10(6)&0.0077(4)&-0.00005(4)&0.972703 \\
\hline
5&1.09(8)&0.0068(6)&-0.00001(4)&0.995759\\
\hline
6&1.1(1)&0.0055(8)&-0.00008(7)&0.869692 \\
\hline
\end{tabular}
\end{center}
\end{table}
\begin{table}[h] 
\begin{center}
\caption{The penetration length $\lambda$ of non-Abelian color electric fields at $\beta = 5.6$ on $24^3 \times 4$ lattices.} \label{lambda_NA}
\begin{tabular}{|c|c|c|c|c|} \hline
\multicolumn{1}{|c|}{$d$} &  \multicolumn{1}{|c|}{$\lambda/a$}&  \multicolumn{1}{|c|}{$c_1$}  &  \multicolumn{1}{|c|}{$c_0$}&$\chi^2/N_{\rm df}$ \\ \hline
3&0.92(2)&0.83(3)&-0.0011(9)& 1.4559\\ 
\hline
4&0.98(6)&0.66(5)&-0.0004(32)&0.866868 \\ 
\hline
5&1.12(6)&0.57(3)&-0.0004(20)&1.21679\\ 
\hline
6&1.23(20)&0.36(6)&-0.0001(43)& 3.13162\\
\hline
\end{tabular}
\end{center}
\end{table}

\subsection{The spatial distribution of color electric fields}
First, we show the results of Abelian color electric fields using an Abelian gauge field $\theta^1_\mu (s)$. To evaluate the Abelian color electric field, we adopt the Abelian plaquette as an operator $O(r)$. We calculate a penetration length $\lambda$ from the Abelian color electric fields for $d=3,4,5,6$ at $\beta=5.6$ and check the $d$ dependence of $\lambda$. To improve the accuracy of the fitting, we evaluate  $O(r)$ at both on-axis and off-axis distances. As a result, we find the Abelian color electric fields $E^A_{z}$ alone are squeezed as in Fig.\ref{EA_zrp}.  We fit these results to a fitting function, 
  \begin{align}
   f(r) = c_1 \mathrm{exp}(-r/\lambda) + c_0 .
  \end{align}  
Here $\lambda$, $c_1$ and $c_0$ are the fit parameters. The parameter $\lambda$ corresponds to the penetration length. Additionally, we calculate the penetration lengths of non-Abelian color electric fields at on-axis to compare them with those of Abelian color electric fields. We find both are almost the same as shown in Table \ref{lambda_A} and \ref{lambda_NA}. We confirm that the penetration length of Abelian color electric fields reproduce the penetration length of non-Abelian color electric fields. 

\subsection{The spatial distribution of monopole currents}
\begin{figure}[h]
\centering
 \hspace*{1cm} 
\includegraphics[keepaspectratio, scale=0.6]{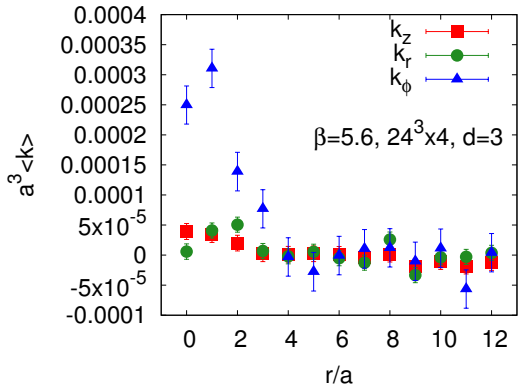} 
\vspace*{.75cm}    
\caption{The profile of monopole current $ k_{\phi}, k_z, k_r$ with $d=3$ at $\beta=5.6$ on $24^3\times 4$ lattices.}\label{mono}    
\end{figure}

Next, we show the result of the spatial distribution of Abelian-like monopole currents.  We define the Abelian-like monopole currents on the lattice as in Eq.(\ref{eq:amon}).
In this study we evaluate the connected correlation (\ref{connect}) between $k^1 (r,\phi,z)$ and two non-Abelian Polyakov loops. 
As a result, we find the spatial distribution of monopole currents around the flux-tube at $\beta = 5.6$. Only the monopole current in the azimuthal direction, $k^1_{\phi}$, shows the correlation with two non-Abelian Polyakov loops as presented in Fig.\ref{mono}.

\subsection{The dual Amp\`{e}re's law}
In previous $SU(2)$ researches \cite{Suzuki:2009xy}, they investigated the dual Amp\`{e}re's law to see what squeezes the color-electric field. In the case of $SU(2)$ gauge theory without gauge fixings, they confirmed the dual Amp\`{e}re's law and the monopole currents squeeze the color-electric fields. In this subsection we show the results of the dual Amp\`{e}re's law in the case of $SU(3)$ gauge theory. The definition of monopole currents gives us  the following relation,
\begin{align}
 (\mathrm{rot}E^{a})_\phi = \partial_{t}B^{a}_{\phi} + 2\pi k^{a}_{\phi}, 
\end{align}
where index $a$ is a color index. 

\begin{figure}[h]
 \centering
  \hspace*{1cm} 
\includegraphics[keepaspectratio, scale=0.6]{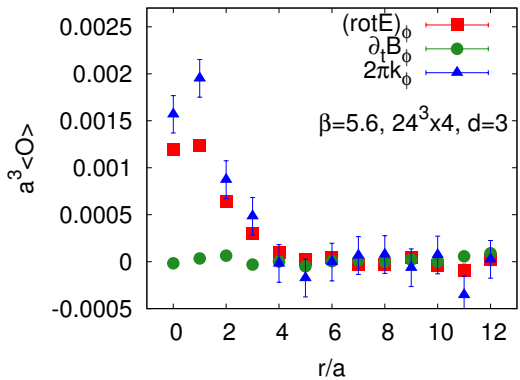} 
\vspace*{.75cm}    
\caption{The dual Amp\`{e}re's law with $d=3$ at $\beta=5.6$ on $24^3\times 4$ lattices.} \label{dualA}
\end{figure}

As a results, we confirm that there is no signal of the magnetic displacement current $\partial_{t}B^{a}_{\phi}$ around the flux-tube for $d=3$ at $\beta=5.6$ as shown in Fig.\ref{dualA}. It suggests that the Abelian-like monopole current squeezes the Abelian color electric field as a solenoidal current in $SU(3)$ gauge theory without gauge fixing, although more data for larger $d$ are necessary.
 

\begin{figure}[h]
\begin{center}
\hspace*{1cm} 
\includegraphics[width=7cm]{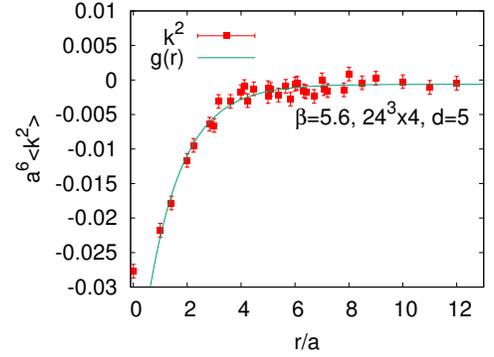}
\end{center}
\vspace*{.5cm} 
\caption{The squared monopole density with $d=5$ at $\beta = 5.6$ on $24^3\times 4$ lattices.}\label{kk}
\end{figure}

\begin{table}[h] 
\begin{center}
\caption{The coherence length $\xi/\sqrt{2}$ at $\beta = 5.6$ on $24^3 \times 4$ lattices.\label{k2}}
\begin{tabular}{|c|c|c|c|c|} \hline
 \multicolumn{1}{|c|}{$d$} &  \multicolumn{1}{|c|}{$\xi/\sqrt{2}a$}&  \multicolumn{1}{|c|}{$c'_1$}  &  \multicolumn{1}{|c|}{$c'_0$}&$\chi^2/N_{\rm df}$ \\ \hline
3&1.04(6)&-0.050(3)& 0.0001(2)&0.997362 \\
\hline
4&1.17(7)&-0.052(3)&-0.0003(2)&1.01499  \\
\hline
5&1.3(1)&-0.047(3)&-0.0006(3)&0.99758 \\
\hline
6&1.1(1)&-0.052(8)&-0.0013(5)&1.12869 \\
\hline
\end{tabular}
\end{center}
\end{table}

\subsection{The vacuum type in $SU(3)$ gauge theory without gauge fixing}
Finally, we evaluate the Ginzburg-Landau parameter, which characterizes the type of the (dual) superconducting vacuum. In the previous result \cite{Suzuki:2009xy}, they found that the vacuum type is near the border between type I and type I\hspace{-.1em}I dual superconductors by using the $SU(2)$ gauge theory without gauge fixing. We apply the same method to $SU(3)$ gauge theory. 

To evaluate the coherence length, we measure the correlation between the squared monopole density and two non-Abelian Polyakov loops by using the disconnected correlation function~\cite{Maxim:2005, Suzuki:2009xy},
\begin{eqnarray}
 \kbra{k^2(r)}_{q\bar{q}}& = &\frac{\kbra{\Tr{P(0)}\Tr{P^{\dagger}(d)}\sum_{\mu, a}k^a_{\mu}(r)k^a_{\mu}(r)}}{\kbra{\Tr{P}(0)\Tr{P^{\dagger}(d)}}} \nonumber \\ 
&& - \kbra{\sum_{\mu,a}k^a_{\mu}(r)k^a_{\mu}(r)}.
\end{eqnarray}
We fit the profiles to the function, 
\begin{eqnarray}
g(r) = c'_1 \mathrm{exp}\left(-\frac{\sqrt{2}r}{\xi}\right) + c'_0,
\end{eqnarray}
where $\xi$, $c'_1$ and $c'_0$ are the fit parameters. The parameter $\xi$ corresponds to the coherence length. We plot the profiles of $\kbra{k^2(r)}_{q\bar{q}} $ in Fig.\ref{kk}.  As a result, we could evaluate the coherence length $\xi$ for $d=3,4,5,6$ at $\beta=5.6$ and find almost the same values of $\xi/\sqrt{2}$ for each $d$ as shown in Table\ref{k2}. Using these parameters $\lambda$ and $\xi$, we could evaluate the Ginzburg-Landau (GL) parameter. The GL parameter  $\kappa = \lambda/\xi$ can be defined as the ratio of the penetration length and the coherence length. If $\sqrt{2}\kappa < 1$, the vacuum is of the type I and if $\sqrt{2}\kappa > 1$, the vacuum  is of the type I\hspace{-.1em}I.  We show the GL parameters in $SU(3)$ gauge theory in Table \ref{GL}. We find that the vacuum  is of the type I near the border between type I and type I\hspace{-.1em}I, although the study is done at one gauge coupling constant $\beta=5.6$. This is the first direct result of the vacuum type in pure $SU(3)$ gauge theory without gauge fixing, although different $\beta$ data are necessary to show the continuum limit. 

\begin{table}[h] 
\begin{center}
\caption{The Ginzburg-Landau parameters at $\beta=5.6$ on $24^3\times 4$ lattice.}\label{GL}
\begin{tabular}{|c|c|} \hline
 \multicolumn{1}{|c|}{d} &  \multicolumn{1}{|c|}{$\sqrt{2}\kappa$} \\ \hline
3&0.87(5) \\
\hline
4&0.93(7)  \\
\hline
5&0.83(9) \\
\hline
6&0.9(2)  \\
\hline
\end{tabular}
\end{center}
\end{table}

\section{Concluding remarks} \label{conclusion}
In this work, we have investigated Abelian dominance, monopole dominance, and the dual Meissner effect in pure $SU(3)$ gauge theory with respect to Abelian-like monopoles without gauge fixing. We have confirmed that these Abelian-like monopoles reproduce the non-Abelian string tension almost perfectly at one gauge coupling constant. And also, we have decided the vacuum type as the type I near the border between type I and I\hspace{-.1em}I by the penetration length from the Abelian color electric fields and the coherence length from the squared monopole density. It is the first Monte-Carlo studies of pure $SU(3)$ QCD with respect to Abelian-like monopoles without any artificial additional assumption such as introduction of partial gauge-fixing.

There are other works~\cite{CCA:PRD86, CCA:PRD89, BB:PRD99} studying the vacuum type in $SU(3)$ QCD, measuring non-Abelian electric fields around static quark pairs. 
Then using a parametrization of the longitudinal component of color electric field around the flux source suggested from the usual superconductor studies, they determine the GL parameter $\kappa$. 
The obtained values of $\kappa$ are different from $0.243(88)$ in \cite{CCA:PRD86}, $0.178(21)$ in \cite{CCA:PRD89} corresponding to Type I to $1.8(6)$~\cite{BB:PRD99} (Type I\hspace{-.1em}I), depending on the method and assumptions adopted. All of them are however indirect contrary to our study here. 

In contrast to our old $SU(2)$ results done in Ref.\cite{Suzuki:2007jp, Suzuki:2009xy}, the $SU(3)$ analyses are unexpectedly hard to get any meaningful results. Especially, we require almost a million vacuum configurations in proving almost perfect monopole dominance. Nevertheless, we get promising results showing our new Abelian-like monopoles play a key role in color confinement also in $SU(3)$ as well as in $SU(2)$. However scaling studies in $SU(3)$ case are not done yet totally. To this purpose, we believe that employing smooth gauge fixings will be helpful to confirm the scaling behavior corresponding to the continuum limit. This is to be done in near future.

\section*{Acknowledgements}
The authors would like to thank Y. Koma for his computer  code of the 
multilevel method. This work used High Performance Computing resources provided by Cybermedia Center of Osaka University through the HPCI System Research Project (Project ID: hp210021). The numerical simulations of this work were done also using High Performance Computing resources at Research Center for Nuclear Physics  of Osaka University, at Cybermedia Center of Osaka University, at Cyberscience Center of Tohoku University and at KEK. The authors would like to thank these centers for their support of computer facilities. T.S was finacially supported by JSPS KAKENHI Grant Number JP19K03848.
\appendix
\section{}
On the lattice, QCD is usually formulated in terms of link fields $U_{\mu}(s)$ as a non-Abelian $SU(3)$ group element. It is not at all trivial to extract Lie-algebra gauge fields $A^a_{\mu}(s)$ for $a=1\sim 8$ from $U_{\mu}(s)$. When we studied $SU(2)$ case in Refs.~\cite{Suzuki:2007jp,Suzuki:2009xy}, we simply extended the method extracting an Abelian gauge field $A^3_{\mu}(s)$ used in MA gauge studies~\cite{Kronfeld:1987vd}  to a case keeping $SU(2)$ gauge symmetry and defined $A^a_{\mu}(s)$ for $a=1\sim 3$.  This can be done, since in $SU(2)$, $U_{\mu}(s)$ is expanded in terms of the Lie-algebra elements as follows:
\begin{eqnarray}
U_{\mu}(s)&=& U^0_{\mu}(s)+i\sum_{a=1}^3 U^a_{\mu}(s)\sigma^a.\label{A1}
\end{eqnarray}
In MAG case, an Abelian link field $\theta^3_{\mu}(s)$ is defined as 
\begin{eqnarray*}
\theta^3_{\mu}(s)&=& \arctan\frac{U^3_{\mu}(s)}{U^0_{\mu}(s)}\ \ (\textrm{mod}\  2\pi).
\end{eqnarray*}
Hence we simply extended this definition to other components having color $a=1$ and $2$ also, since without any partial gauge-fixing like MAG, $SU(2)$ symmetry is not broken. This definition works very well as seen from the numerical results obtained in Refs.~\cite{Suzuki:2007jp,Suzuki:2009xy,Suzuki:2017lco,Suzuki:2017zdh}

However in $SU(3)$, the situation is completely different. To get a relation like Eq.(\ref{A1}), we first diagonalize $U_{\mu}(s)$ by a unitary matrix $V(s)$. Then we get 
\begin{eqnarray*}
U_{\mu}(s)=V(s)\left(
  \begin{array}{ccc}
    \Lambda^1_{\mu}(s)   & 0   & 0   \\
      0 & \Lambda^2_{\mu}(s)   & 0   \\
      0 &  0  &\Lambda^3_{\mu}(s)    \\
  \end{array}
\right)
V^{\dag}(s).
\end{eqnarray*}
Since the diagonal part can be written in terms of $3\times 3$ unit matrix and diagonal Gell-Mann matrices $\lambda_3$ and $\lambda_8$. Formally we can get a relation like (\ref{A1}) in $SU(3)$, but the coefficients of the Gell-Mann matrices $\lambda_a$ are not real in general.  Hence we can not adopt the same simple definition as done in (\ref{A1}). But here it is interesting to note that the same definition (\ref{A1}) in $SU(2)$ can be obtained also by maximizing the norm
\begin{eqnarray*}
RA= \mathrm{Re} \tr\left\{\exp(i\theta_\mu^a(s)\sigma^a)U_\mu^{\dag}(s)\right\},
\end{eqnarray*}
as done in Eq.(\ref{RA}). This definition can be extended easily to $SU(3)$ as adopted here in Eq.(\ref{RA}).

\end{document}